\documentclass[twocolumn]{aastex631}

\usepackage{amsfonts}
\usepackage{amsmath}
\usepackage{amssymb}
\usepackage{subfigure}
\usepackage{graphicx}
\usepackage{epstopdf}
\usepackage{epsfig}
\usepackage{float}
\usepackage{hyperref}
\usepackage{color}
\usepackage{ulem}
\usepackage{cancel}
\usepackage{mathrsfs}
\usepackage{booktabs}
\usepackage{multirow}
\usepackage{natbib}

\usepackage{xcolor}

\begin{document}

\title{The Extreme Faraday Effect in Fast Radio Bursts}

\author[0009-0009-7749-8998]{Yu-Chen Huang}
\affiliation{Department of Astronomy, University of Science and Technology of China, Hefei 230026, China; daizg@ustc.edu.cn}
\affiliation{School of Astronomy and Space Science, University of Science and Technology of China, Hefei 230026, China}

\author[0000-0002-7835-8585]{Zi-Gao Dai}
\affiliation{Department of Astronomy, University of Science and Technology of China, Hefei 230026, China; daizg@ustc.edu.cn}
\affiliation{School of Astronomy and Space Science, University of Science and Technology of China, Hefei 230026, China}

\begin{abstract}

Fast radio bursts (FRBs) are a type of highly-polarized, millisecond-duration electromagnetic pulses in the radio band, which are mostly produced at cosmological distances. These properties provide a natural laboratory for testing the extreme Faraday effect, a phenomenon in which two different propagation modes of a pulse separate after passing through a dense, highly ionized, and magnetized medium. We derive the critical condition (e.g., rotation measure) for the extreme Faraday effect to occur in FRBs, which exceeds the currently observed maximum value but remains within the theoretically predicted range. Some new features of FRBs (in particular, radio bursts with much shorter durations) after undergoing the extreme Faraday effect are predicted, such as sudden sign reversals of circular polarization, conspicuous frequency drifting, and emergency of extremely high circular polarization degrees. A potential application of this effect in FRBs is that, by comparing morphological differences of the two separated twin modes, one can identify the variations of plasma properties over extremely short timescales along the propagation path. Therefore, if this effect is found with future observations, it would provide a new tool for probing dense, magnetized environments near FRB sources.

\end{abstract}


\section{Introduction}

Fast radio bursts (FRBs) are a class of radio transients with unknown physical origins, typically from cosmic distances \citep{Cordes2019,Petroff2019,Xiao2021,Zhang2023,Wu2024a}. Their short timescales ($\sim\text{ms}$), low frequencies ($\sim\text{GHz}$), and high linear polarization make them an unparalleled tool for probing the magneto-ionic environments close to their sources \citep{Masui2015,Michilli2018,Feng2022,Xu2022,AnnaThomas2023,Li2025}. A quantity called rotation measure (RM), defined by
\begin{equation}
	\text{RM}=\frac{e^3}{2\pi m_e^2 c^4}\int n_e B_{\parallel}dl,
	\label{rm}
\end{equation}
can be used to quantify the integrated magnetic field strength along the line of sight. In Eq. (\ref{rm}), $e$,  $m_e$, and $n_e$ are the electron charge, mass, and number density, respectively. $c$ is the speed of light, and $B_{\parallel}$ is the magnetic field strength along the line of sight. To obtain RM, a standard approach is to measure the polarization angles (PAs) $\phi$ of different wavelengths $\lambda$ of a linearly polarized FRB, and then make use of the relation
\begin{equation}
	\phi\simeq\text{RM}\lambda^2,
	\label{philambda2}
\end{equation}
which is derived from the Faraday effect.

\begin{figure}
	\begin{center}
		\subfigure{
			\includegraphics[width=0.45\textwidth]{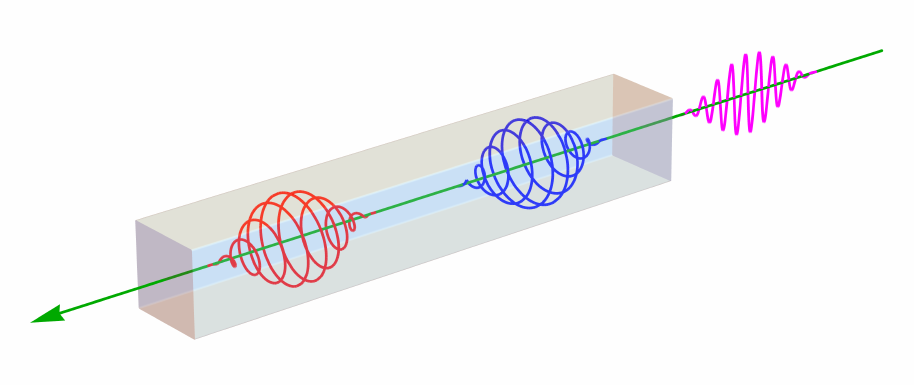}}
		\caption{Schematic illustration of the extreme Faraday effect. An incident electromagnetic pulse (magenta) is linearly polarized and propagates along a background magnetic field (green arrow). When the RM exceeds the critical value given by Eq. (\ref{rmcri}), the pulse splits into two circularly polarized sub-pulses (red and blue) due to different group velocities between the two sub-pulses.}
		\label{splittingrlmodes}
	\end{center}
\end{figure}

We suggest that this conventional method is limited by what is known as the extreme Faraday effect \citep{Weng2017}, which sets a measurable maximum critical value,
\begin{equation}
	\text{RM}_{c}\simeq\frac{\omega^3 W}{16\pi^2 c^2}\simeq\left(1.7\times10^{7}\text{ rad}\text{ m}^{-2}\right)\nu_{\text{GHz}}^3 W_{-3},
	\label{rmcri}
\end{equation}
where $\nu=\omega/2\pi=\left(1\text{ GHz}\right)\nu_{\text{GHz}}$ and $W=\left(10^{-3}\text{ s}\right)W_{-3}$ are the typical frequency and intrinsic temporal width of a radio burst, respectively. This equation is derived below and a general discussion on it is seen in Appendix \ref{gdes}. Beyond the critical RM, a linearly polarized burst splits into two circularly polarized sub-bursts that propagate independently, as illustrated in Fig. \ref{splittingrlmodes}. So far, the maximum RM detected in FRBs is $\sim10^5\text{ rad m}^{-2}$ \citep{Michilli2018}, which is two orders of magnitude lower than Eq. (\ref{rmcri}). Nevertheless, an even larger RM, on the order of $\sim10^7$ to $\sim10^8\text{ rad m}^{-2}$ is possible, such as when the progenitor of an FRB is embedded in an early-stage supernova remnant \citep{Piro2016,Zhao2021,Zhao2021a} or near a supermassive black hole \citep{Li2023,Yang2023,Zhao2024}.

Another way to reach the critical RM is to reduce the frequency of observation. \cite{Suresh2019} studied the similar magnetic splitting effect and proposed that it could leave a detectable imprint in the low-frequency band (several hundred MHz). However, due to strong scattering smearing, the imprint deviates significantly from the intrinsic splitting behavior. Recently, more and more FRBs with timescales down to microseconds or even nanoseconds have been identified \citep{Farah2018,Cho2020,Majid2021,Nimmo2021,Nimmo2022,Snelders2023,Sand2025}. These findings may indicate that some FRBs possess intrinsic extremely narrow widths and suggest the feasibility of observing significant splitting at relatively high frequencies in these FRBs.

In this paper, we aim to investigate the region beyond the critical value given by Eq. (\ref{rmcri}). In this region, a series of new features emerge in FRBs, especially in those with narrow widths, which could in principle be detected in future observations. We present a model of the extreme Faraday effect in a magnetized electron--ion plasma in Section \ref{model}, and then generalize the model to an electron--positron pair plasma with a strong background magnetic field in Section \ref{pairplasma}. We propose a criterion for telescopes to observe the effect in Section \ref{telescope}, and investigate the extreme Faraday effect in short-timescale FRBs in Section \ref{stimescales}. We discuss some observational challenges in Section \ref{ochallenges}, where we also identify the optimal frequency for detecting the effect in FRBs. The conclusions and discussions are summarized in Section \ref{cdiscussion}. The notation $Q_n=Q/10^n$ and cgs units are adopted in this paper.

\section{Model}\label{model}

In a cold, magnetized electron--ion plasma, there are two allowed propagation modes for an electromagnetic wave: the left-handed circularly polarized (L-) mode and the right-handed circularly polarized (R-) mode. We first consider a simple case when the propagation direction of the wave is parallel to the background magnetic field. These two modes then have dispersion relations \citep{Boyd2003}
\begin{equation}
	\begin{aligned}
		\frac{c^2 k^2}{\omega^2}&\simeq1-\frac{\omega_p^2}{\omega\left(\omega+\omega_B\right)},&\text{L-mode},\\
		\frac{c^2 k^2}{\omega^2}&\simeq1-\frac{\omega_p^2}{\omega\left(\omega-\omega_B\right)},&\text{R-mode}.
	\end{aligned}\label{disperrl}
\end{equation}
In the above equation, $k$ is the wave vector, $\omega_p^2=4\pi e^2\left(n_i/m_i+n_e/m_e\right)\simeq4\pi n_e e^2/m_e$ is the square of plasma frequency, $n_i$ is the ion number density, $m_i$ is the ion mass, and $\omega_B=eB_{\parallel}/m_e c$ is the electron cyclotron frequency in a background field $B_{\parallel}$. According to the dispersion relation, one can derive the phase velocity $v_p$ and the group velocity $v_g$ of these two modes. In most astrophysical environments with electron--ion plasmas, the plasma number density and magnetic field strength usually deviate significantly from their critical values, which are given by $n_c=m_e \omega^2/4\pi e^2\simeq\left(1.2\times10^{10} \text{ cm}^{-3}\right)\nu_\text{GHz}^{-2}$ and $B_{\parallel,c}=m_e c \omega/e\simeq\left(3.6\times10^2\text{ G}\right)\nu_\text{GHz}$. We thus focus on the case in which $\omega_p\ll\omega$ and $\omega_B\ll\omega$ are satisfied. The phase velocity difference and the group velocity difference between the L-mode and the R-mode then reduce to
\begin{equation}
	\begin{aligned}
		\frac{\Delta v_p}{c}&=\frac{v_{p,L}-v_{p,R}}{c}\simeq-\left(\frac{\omega_p}{\omega}\right)^2 \left(\frac{\omega_B}{\omega}\right),\\
		\frac{\Delta v_g}{c}&=\frac{v_{g,L}-v_{g,R}}{c}\simeq 2\left(\frac{\omega_p}{\omega}\right)^2 \left(\frac{\omega_B}{\omega}\right).
	\end{aligned}\label{deltavgei}
\end{equation}
An electromagnetic wave propagating in the plasma can be regarded as the superposition of these two independent propagation modes.

On the one hand, the difference between the phase velocities of these two circularly polarized modes can lead to a rotating PA of the original linearly polarized electromagnetic wave. The PA can be written as
\begin{equation}
	\phi=\frac{1}{2}\int\left(k_L-k_R\right)dl\simeq\frac{e^3 \lambda^2}{2\pi m_e^2 c^4}\int n_e B_{\parallel}dl,
\end{equation}
which is Eq. (\ref{philambda2}). On the other hand, the different group velocities of both modes indicate that they tend to separate during propagation. When the R-mode wave is delayed by time $W$ relative to the L-mode wave, i.e.,
\begin{equation}
	\int\frac{dl}{v_{g,R}}-\int\frac{dl}{v_{g,L}}=W,
\end{equation}
the splitting between these two modes is distinguishable. Using Eq. (\ref{deltavgei}), one obtains
\begin{equation}
	\int n_e B_{\parallel}dl\simeq\frac{m_e^2 c^2 \omega^3 W}{8\pi e^3}.
	\label{nbdlei}
\end{equation}
Substituting Eq. (\ref{nbdlei}) into Eq. (\ref{rm}) results in Eq. (\ref{rmcri}).\footnote{A similar equation of the required RM for the extreme Faraday effect was presented by Su-Ming Weng at \href{https://frbconference.scievent.com/home/}{the Second China FRB Symposium}.} This sets a maximum RM that is measurable conventionally and a critical condition for the extreme Faraday effect to occur in FRBs. For a general angle between the propagation direction and the background magnetic field, these equations remain valid after replacing the background magnetic field with its component parallel to the propagation direction.

\begin{figure*}
	\begin{center}
		\subfigure{
			\includegraphics[width=0.32\textwidth]{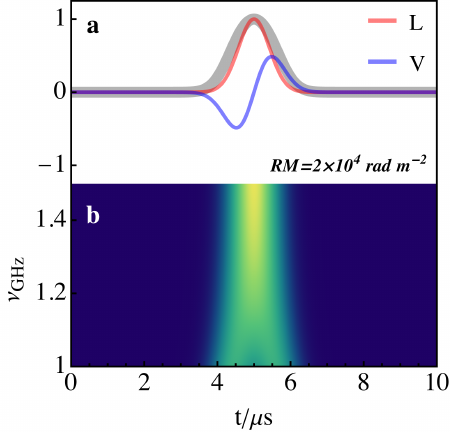}}
		\subfigure{
			\includegraphics[width=0.32\textwidth]{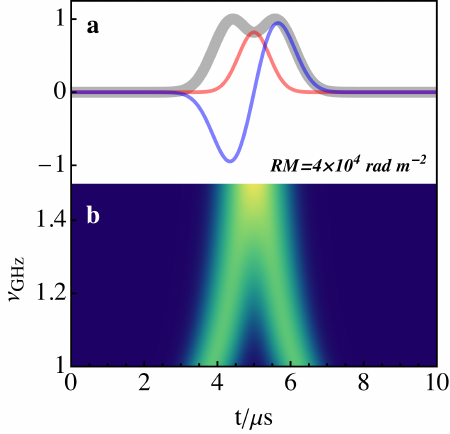}}
		\subfigure{
			\includegraphics[width=0.32\textwidth]{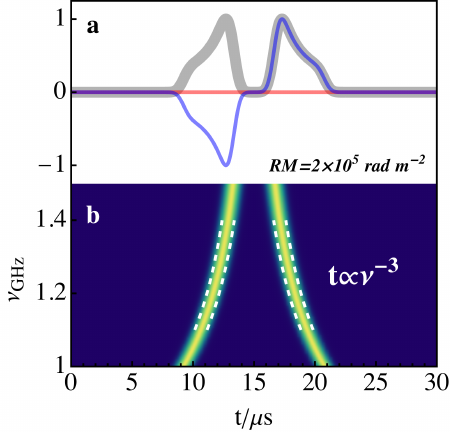}}
		\caption{Theoretical prediction of (\textbf{a}) the burst profile and (\textbf{b}) the dynamic spectrum of a microsecond-duration radio burst after passing through a dense magnetized electron--ion plasma. The gray thick curve, the red curve, and the blue curve represent the frequency-averaged intensity $I$, linearly polarized intensity $L\equiv\sqrt{Q^2 + U^2}$, and circularly polarized intensity $V$, respectively. The only varying parameter in these panels is RM, with adopted values $2\times 10^4$, $4\times10^4$, and $2\times10^5\text{ rad m}^{-2}$ from left to right. The dynamic spectra of the L-mode sub-pulse and the R-mode sub-pulse are symmetric about the arrival time, both showing conspicuous frequency drifting with $t\propto\nu^{-3}$ profiles, which are outlined by white dished curves. The original radio burst is assumed to be fully linearly polarized and has no intrinsic frequency drifting.}
		\label{dyspecstokesei}
	\end{center}
\end{figure*}

FRBs experience strong dispersion when propagating in plasmas, which indicates that components with different frequencies arrive at different times (e.g., $\nu_\star$ and $\nu_\dagger$), i.e., 
\begin{equation}
	\begin{aligned}
		\Delta t&=t\left(\nu_\star\right)-t\left(\nu_\dagger\right)\\
		&\simeq\left(4.1\times10^{-3}\text{ s}\right)\left(\nu_{\star,\text{GHz}}^{-2}-\nu_{\dagger,\text{GHz}}^{-2}\right)\text{DM}\\
		&\pm\left(2.9\times10^{-11}\text{ s}\right)\left(\nu_{\star,\text{GHz}}^{-3}-\nu_{\dagger,\text{GHz}}^{-3}\right)\text{RM},
	\end{aligned}\label{deltat}
\end{equation}
where the positive or negative sign refers to the R-mode or L-mode. In the above equation, RM is in units of rad m$^{-2}$, and  dispersion measure (DM) is in units of pc cm$^{-3}$. DM is defined as the integral of the electron number density along the line of sight,
\begin{equation}
	\text{DM}=\int n_e dl.
	\label{dm}
\end{equation}
The DM term completely dominates the arrival time in Eq. (\ref{deltat}), unless the averaged magnetic field strength along the line of sight $\langle B_{\parallel}\rangle\gtrsim100\text{ G}$, which is unrealistic. However, the contribution of the DM term can be eliminated by fitting the $t\propto\nu^{-2}$ curve in the frequency-arrival time plot. By assuming some original parameters, one can predict the dynamic spectrum, pulse profile, and the polarization state of an FRB after passing through a magnetized, dense plasma, as shown in Fig. \ref{dyspecstokesei}.

In the extreme Faraday effect, the interval between the two sub-pulses is determined by the RM. For a relatively small RM, the maximum degree of linear polarization always appears in the overlap region of the two sub-pulses. In contrast, the degree of circular polarization peaks at each sub-pulse, and exhibits a sign-switching behavior from pulse to pulse. For a relatively large RM, the original pulse completely splits into two circularly polarized sub-pulses. The absence of linearly polarized components indicates that RM cannot be derived from the traditional $\phi=\text{RM}\lambda^2$ relation. Consequently, a novel approach can be developed to obtain the RM value, which is to fit the frequency drifting curve $t\propto\nu^{-3}$ in the dynamic spectrum by excluding the DM term in Eq. (\ref{deltat}).

Theoretically, similar sub-pulses may originate from other propagation effects, such as multipath propagation, dispersion, or scattering. Moreover, various frequency drifting patterns are commonly observed in the dynamic spectra of FRBs \citep{Hessels2019,Pleunis2021,Zhou2022}, which may be attributed to intrinsic radiation mechanisms \citep{Wang2019}. In practical observations, these phenomena may sometimes be confused with the extreme Faraday effect. We therefore conclude several distinct features of the extreme Faraday effect to distinguish it from other phenomena: (1) The extreme Faraday effect occurs only when the RM reaches or exceeds the critical value. In contrast, identifying whether sub-pulses originate from most of the aforementioned propagation effects may be ambiguous due to various uncertain parameters. (2) The frequency drifting in the extreme Faraday effect always appears in pairs in the dynamic spectrum, with the earlier sub-pulse drifting upward and the later sub-pulse drifting downward. In comparison, downward drifting predominates in common frequency drifting \citep{Zhou2022}. (3) The two separated sub-pulses with different drifting directions exhibit high degrees of circular polarization (can reach as high as $100\%$ ideally) with opposite signs. This is especially useful to identify the extreme Faraday effect if polarization information is available. (4) If many bursts are detected from a repeater source, the RM of the source can be measured by both frequency drifting and the conventional method, with the latter being more effective for bursts with higher frequencies or longer durations, which exhibit strong linear polarization components (i.e., weak extreme Faraday effects). If RM variations between bursts can be neglected, we can compare the RM values measured by the two methods to further confirm the extreme Faraday effect.

\section{Pair Plasma}\label{pairplasma}

\begin{figure}
	\begin{center}
		\subfigure{
			\includegraphics[width=0.45\textwidth]{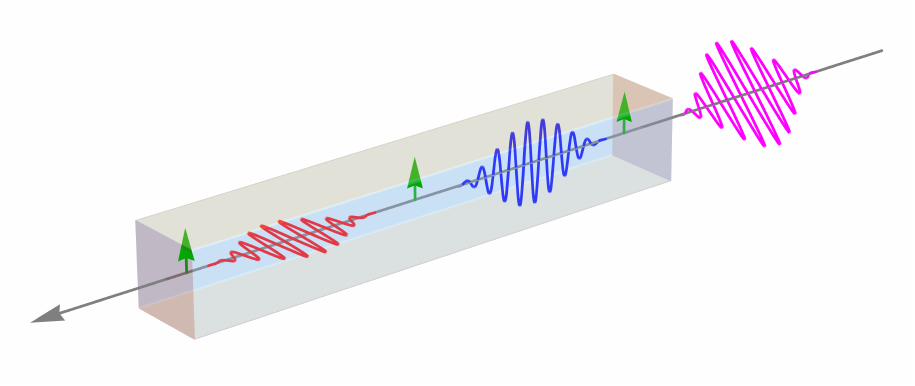}}
		\caption{Schematic illustration of the extreme Faraday effect in an electron--positron pair plasma. The propagation direction (gray arrow) of the original linearly polarized pulse (magenta) is assumed to be perpendicular to the background magnetic field (green arrows). The pulse ultimately splits into two linearly polarized sub-pulses (red and blue) with orthogonal polarization directions.}
		\label{splittingxomodes}
	\end{center}
\end{figure}

The extreme Faraday effect can be generalized to an electron--positron pair plasma, which is likely an important component around FRB sources. An electromagnetic wave propagating in a pair plasma behaves quite differently from that in an electron–ion plasma. When the propagation direction of the wave is parallel to the background magnetic field, electrons and positrons contribute equal circular polarization to the wave. The Faraday effect thus disappears in a quasi-neutral pair plasma.\footnote{If a radio wave propagates through a pair plasma, where electrons and positrons have different distributions, its state of polarization could be altered, see \cite{Wang2010} for a discussion.} A non-trivial case occurs when the propagation direction is oblique to the background field. For simplicity, we first consider the case of perpendicular propagation. The dispersion relations of two permitted propagation modes in a pair plasma are given by \citep{Zhang2023}
\begin{equation}
	\begin{aligned}
		\frac{c^2 k^2}{\omega^2}&\simeq1-\frac{\omega_p^2}{\omega^2-\omega_B^2},&\text{X-mode},\\
		\frac{c^2 k^2}{\omega^2}&\simeq1-\frac{\omega_p^2}{\omega^2},&\text{O-mode},
	\end{aligned}
\end{equation}
where the letters X- and O- denote the extraordinary and ordinary modes, respectively. Both modes are linearly polarized but have orthogonal polarization directions. The electric field of the X-mode wave is perpendicular to the plane defined by the wave vector and the background magnetic field, while the electric field of the O-mode wave lies in the plane. The different group velocities of these two modes also indicate a tendency of separation during propagation. A schematic illustration of the extreme Faraday effect in a pair plasma is shown in Fig. \ref{splittingxomodes}.

In astrophysics, an electron--positron pair plasma likely exists in the region with high energy density, such as the magnetosphere of a neutron star \citep{Goldreich1969}. The magnetic field in the stellar magnetosphere is so strong that the conditions $\omega\ll\omega_B$ and $\omega_p\ll\omega_B$ are usually satisfied. Therefore, the difference between group velocities of the X-mode and O-mode waves reduces to
\begin{equation}
	\frac{\Delta v_{g,\text{pair}}}{c}=\frac{v_{g,X}-v_{g,O}}{c}\simeq1-\left(1-\frac{\omega_p^2}{\omega^2}\right)^{1/2}.
	\label{deltavgep}
\end{equation}
A critical condition for the extreme Faraday effect to occur in a pair plasma is
\begin{equation}
	\int_{\text{pair}}\frac{dl}{v_{g,O}}-\int_{\text{pair}}\frac{dl}{v_{g,X}}=W.
\end{equation}
Using Eq. (\ref{deltavgep}), one finally arrives at
\begin{equation}
	\begin{aligned}
		\text{DM}_{\text{pair},c}&=\int_{\text{pair}}  n_e dl\simeq\frac{m_e c \omega^2 W}{4\pi e^2}\\
		&\simeq \left(1.2\times10^{-4}\text{ pc cm}^{-3}\right)\nu_\text{GHz}^2 W_{-6},
	\end{aligned}\label{epnedl}
\end{equation}
where $\text{DM}_{\text{pair}}$ is the dispersion measure in a pair plasma. We have also assumed that the particle number density $n\ll n_c$ (e.g., $\omega_p^2 \ll\omega^2$) for simplicity.\footnote{Another consideration is that according to the dispersion equation, if $\omega_p^2 \ll\omega^2$ is violated, the propagation direction of the O-mode wave will deviate from the direction of the wave vector, which may make the O-mode wave undetectable, see \cite{Lu2019} for a discussion.} For the case of oblique propagation, the integral term in Eq. (\ref{epnedl}) should be generalized to $\int_{\text{pair}}\left(\sin{\theta}\right)^2 n_e dl$, where $\theta$ is the angle between the propagation direction and the background magnetic field. This equation provides a threshold on $\text{DM}_{\text{pair}}$ for the extreme Faraday effect in a pair plasma.

\begin{figure*}
	\begin{center}
		\subfigure{
			\includegraphics[width=0.32\textwidth]{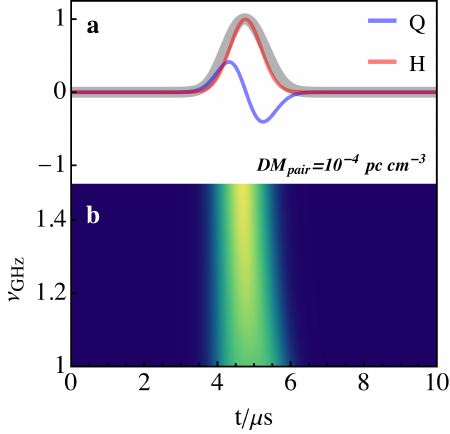}}
		\subfigure{
			\includegraphics[width=0.32\textwidth]{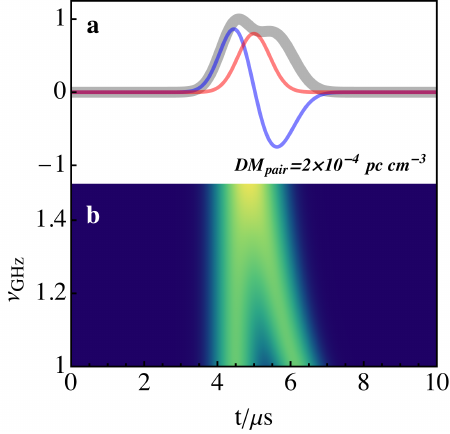}}
		\subfigure{
			\includegraphics[width=0.32\textwidth]{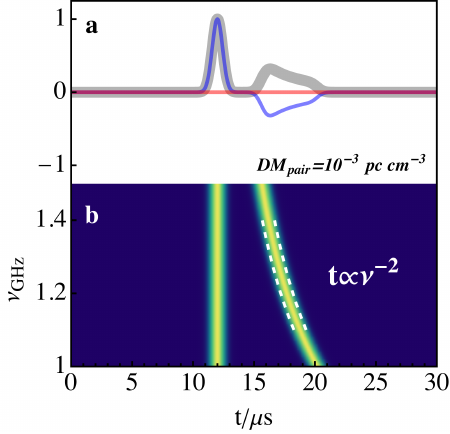}}
		\caption{Theoretical prediction of (\textbf{a}) the burst profile and (\textbf{b}) the dynamic spectrum of a microsecond-duration radio burst after passing through a dense magnetized electron--positron pair plasma. The gray thick curve, the blue curve, and the red curve represent the frequency-averaged intensity $I$, Stokes parameter $Q$, and parameter $H\equiv\sqrt{U^2+V^2}$, respectively. The only varying parameter in these panels is DM$_{\text{pair}}$, with adopted values $10^{-4}$, $2\times10^{-4}$, and $10^{-3}\text{ pc cm}^{-3}$ from left to right. An abrupt orthogonal jump in the PA appears when the two sub-pulses are well separated. Only the O-mode branch shows frequency drifting, which can be characterized by $t\propto\nu^{-2}$. The original pulse is assumed to be fully linearly polarized, and has equal X-mode and O-mode components. The intrinsic frequency drifting of the original pulse is omitted.}
		\label{dyspecstokesep}
	\end{center}
\end{figure*}

The difference in arrival times between different-frequency components can be written as
\begin{equation}
	\begin{aligned}
		\Delta t_{\text{pair}}=&t_{\text{pair}}\left(\nu_\star\right)-t_{\text{pair}}\left(\nu_\dagger\right)\\
		\simeq&\left(4.1\times10^{-3}\text{ s}\right)\left(\nu_{\star,\text{GHz}}^{-2}-\nu_{\dagger,\text{GHz}}^{-2}\right)\\
		&\times\left(\text{DM}+2\text{DM}_{\text{pair}}\right).
	\end{aligned}
\end{equation}
Note that the term DM$_{\text{pair}}$ vanishes for the X-mode wave, since its dispersion in a pair plasma is negligible. This means that only the O-mode component of a radio burst exhibits frequency drifting. The theoretically predicted  pulse profile is exhibited in Fig. \ref{dyspecstokesep}. In the pair plasma, the difference in phase velocities between the X-mode and O-mode waves leads to rapid Faraday conversion between linear polarization and circular polarization components. In this case, the linear polarization degree $L$ and circular polarization degree $V$ both oscillate rapidly with frequency. Therefore, we use the frequency-independent Stokes parameters $Q$ and $H\equiv\sqrt{U^2+V^2}$ to describe the global polarization behavior of the pulse during the splitting process.

A larger $\text{DM}_{\text{pair}}$ usually corresponds to a wider interval between the two sub-pulses. For a small $\text{DM}_{\text{pair}}$, the Stokes parameter $Q$ exhibits a sign reversal behavior, which is similar to the Stokes parameter $V$ in an electron--ion plasma. When $\text{DM}_{\text{pair}}$ is large enough, the original pulse completely splits into two sub-pulses, and an abrupt orthogonal jump in the PA between the two sub-pulses occurs. In this case, the DM value can be obtained in principle by comparing the arrival time difference between the X-mode and O-mode branches in the dynamic spectrum.

One possible scenario for the extreme Faraday effect in an electron--positron pair plasma is in a binary system, where the companion of the FRB source is a neutron star. An order of magnitude estimate on the contribution of magnetospheric electron--positron pairs of the companion to dispersion is
\begin{equation}
	\begin{aligned}
		\text{DM}_{\text{pair}}&\sim\mathcal{M}n_{\text{GJ}}R\\
		&\sim\left(2.3\times10^{-5}\text{ pc cm}^{-3}\right)\mathcal{M}B_{s,12}P_{-1}^{-1}R_8^{-2},
	\end{aligned}
	\label{dmestimate}
\end{equation}
where $\mathcal{M}$ is a multiplicity factor, and $n_{\text{GJ}}$ is the Goldreich-Julian density \citep{Goldreich1969}. Furthermore, $R$, $B_s=\left(10^{12}\text{ G}\right)B_{s,12}$, and $P$ are the magnetospheric radius, surface magnetic field strength, and spin period of the companion. One can see that Eq. (\ref{dmestimate}) is comparable to the critical value in Eq. (\ref{epnedl}) when
\begin{equation}
	W\sim\left(1.9\times10^{-7}\text{ s}\right)\mathcal{M}B_{s,12}P_{-1}^{-1}R_8^{-2}\nu_\text{GHz}^{-2}.
\end{equation}
The value of $\mathcal{M}$ should satisfy
\begin{equation}
	\mathcal{M}\ll\frac{n_c}{n_{\text{GJ}}}\sim1.7\times10^4 B_{s,12}^{-1}P_{-1}R_8^3 \nu_\text{GHz}^2.
\end{equation}
One remaining problem is that the probability of the line of sight passing through the magnetosphere of a neutron star companion may be low, but it is not impossible.

\section{A Criterion for Telescopes}\label{telescope}

The maximal RM that a telescope is capable of detecting by the conventional method (\ref{philambda2}) may be limited by its frequency resolution. The polarization state of an FRB can be well characterized only when the oscillation of Stokes parameters is distinguishable in the spectrum. This naturally sets a requirement for the frequency resolution of the instrument. One can assume that the frequency resolution of the telescope is $\delta\nu$. Then the wavelength resolution is $\delta\lambda=\left(\lambda/\nu\right)\delta\nu$. The resolution of $\lambda^2$ thus is $\delta(\lambda^2)\sim\lambda\delta\lambda\sim\left(\lambda^2/\nu\right)\delta\nu$. According to Eq. (\ref{philambda2}), the maximal RM that is measurable by the instrument should be
\begin{equation}
	\text{RM}_{\text{ins}}\sim\frac{1}{\lambda^2}\left(\frac{\delta\nu}{\nu}\right)^{-1},
\end{equation}
which is supposed to exceed the critical value RM$_c$ in Eq. (\ref{rmcri}). This yields 
\begin{equation}
	W\delta\nu<1.
	\label{criterion}
\end{equation}
An instrument with a frequency resolution that violates the inequality will detect bursts that are significantly depolarized. The criterion is also applicable in a pair plasma. For a microsecond FRB, the required frequency resolution $\delta\nu<\left(1\text{ MHz}\right)W_{-6}^{-1}$. Notably, this requirement is not mandatory for measuring the Stokes parameter V in an electron--ion plasma, the Stokes parameter Q in an electron--positron pair plasma, or when the two sub-pulses are completely separated.

\section{FRBs with Shorter Timescales}\label{stimescales}

\begin{figure}
	\begin{center}
		\subfigure{
			\includegraphics[width=0.45\textwidth]{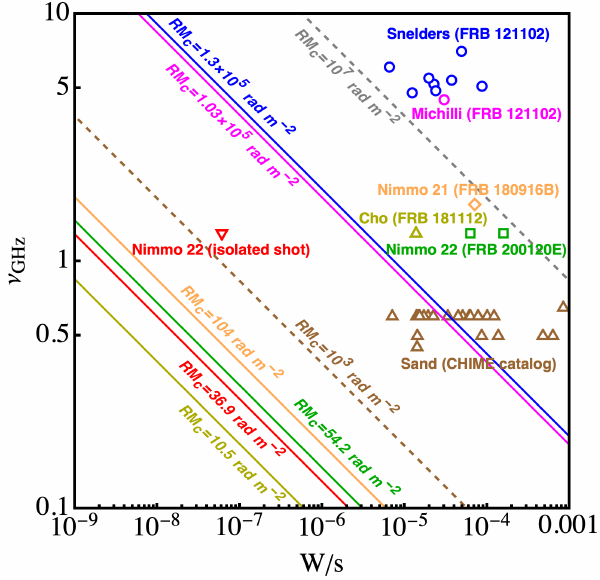}}
		\caption{Frequency-width diagram of a few selected short-duration FRBs. Bursts from different papers are distinguished by different colors and shapes. Lines of different colors represent the cases in which the observed RM values of these bursts equal their critical values given by Eq. (\ref{rmcri}). The gray dashed line represents $\text{RM}_c=10^7\text{ rad m}^{-2}$. In the selected CHIME sample (brown triangles), the component with the minimum width in each burst is adopted. The brown dashed line represents $\text{RM}_c=10^3\text{ rad m}^{-2}$, which is almost the maximum RM within the CHIME sample. The central frequency is adopted for these bursts. All bursts lie above their corresponding colored lines, which suggests that the conditions for the extreme Faraday effect are not met in these observations.}
		\label{nuw}
	\end{center}
\end{figure}

Some bursts with extremely short timescales have already attracted wide interest \citep{Farah2018,Cho2020,Majid2021,Nimmo2021,Nimmo2022,Snelders2023,Sand2025}. These bursts can provide valuable clues for seeking the extreme Faraday effect. Some samples selected from these papers are plotted in Fig. \ref{nuw}, which shows that their timescales and frequencies are still too high to trigger pronounced splitting. We notice that the intrinsic temporal width of a burst is likely independent of its RM, which is mainly contributed by its surrounding environment. If the bursts with the shortest timescale are detected in the sources with the largest RM, the extreme Faraday effect is still likely to occur. Therefore, observations with high time resolution toward FRB sources with large RMs and searching for sources with larger RMs are still called for.

\section{Observational Challenges}\label{ochallenges}

\subsection{Free-Free Absorption}

A large RM normally indicates a dense magneto-ionic near-source environment, where the free-free absorption of the ionized medium may impede the propagation of FRBs. The free-free absorption coefficient in the Rayleigh-Jeans regime reads \citep{Rybicki1986}
\begin{equation}
	\alpha_\nu^{\text{ff}}\simeq1.8\times10^{-2}T^{-3/2}Z^2  \nu^{-2}n_e n_i\bar{g}_{\text{ff}},
\end{equation}
where $T$ is the temperature of the ionized medium in units of $K$, $Z$ is the charge per ion, and $\bar{g}_{\text{ff}}$ is the averaged Gaunt factor. For an ionized medium with a characteristic length scale $L$, a radio wave is transparent when the free-free absorption optical depth $\alpha^{\text{ff}}_\nu L\lesssim1$. If the progenitor of an FRB is embedded in a supernova remnant, one could expect that the maximum RM permitted by free-free absorption is
\begin{equation}
	\text{RM}_{\text{ff}}\sim\left(2\times10^7\text{ rad m}^{-2}\right)L_{17}^{1/2}T_6^{3/4}B_{-3}\nu_{\text{GHz}},
	\label{confreefree}
\end{equation}
where we have adopted $Z\approx1$, $\bar{g}_{\text{ff}}\approx1$, and $n_i\approx n_e$ for simplicity. The parameters adopted above are derived from those expected in young supernova remnants at the age of several decades \citep{Piro2016}. Similarly, if the process occurs in the disk of an active galactic nucleus, the ejecta could contribute a even larger RM after the environment becomes transparent \citep{Zhao2024}. If the free-free absorption is too strong, FRBs will not be able to escape or be detected.

\subsection{Depolarization due to Multipath Propagation}

\cite{Feng2022} reported that repeaters with high RMs tend to exhibit decreasing degrees of polarization at lower frequencies. This is interpreted by multipath propagation of rays with different RM values. One can define a critical frequency
\begin{equation}
	\nu_{\text{dep}}\sim c\sigma_{\text{RM}}^{1/2},
	\label{nudepolarization}
\end{equation}
below which the depolarization comes into play. The quantity $\sigma_{\text{RM}}$ in units of $\text{rad m}^{-2}$ is the dispersion of RM. For FRB 121102A, $\sigma_{\text{RM}}\approx30.9\text{ rad m}^{-2}$, which gives a depolarization frequency $\sim1.7\text{ GHz}$ \citep{Feng2022}. Importantly, the depolarization due to multipath propagation only affects the degree of linear polarization. This is easy to understand because the superposition of rays with different RMs does not reduce the total degree of circular polarization (see Appendix \ref{depefe} for a discussion). Therefore, depolarization due to multipath propagation does not have a significant impact on the extreme Faraday effect since the variation of circular polarization is a more important feature.

\subsection{Scattering}

When scattering-induced temporal broadening dominates the pulse profile, the signal of the extreme Faraday effect may be smeared. Assuming the scattering timescale follows a simple power-law relation with frequency $\tau_{\text{sc}}\propto\nu^{-4}$, one can quantify the degree of scattering by evaluating a dimensionless and relative scattering broadening $\widetilde{\tau}_{\text{sc}}$, which is defined as the scattering broadening divided by the temporal interval between the two sub-pulses,
\begin{equation}
	\widetilde{\tau}_{\text{sc}}\simeq1.7\times10^{10}\tau^*_{\text{sc}}\text{RM}^{-1}\nu_{\text{GHz}}^{-1},
	\label{eta}
\end{equation}
where $\tau_\text{sc}^{*}$ is the scattering timescale in units of second at 1 GHz frequency. In particular, the splitting becomes more pronounced as $\widetilde{\tau}_{\text{sc}}$ decreases. The impact of scattering is insignificant when $\widetilde{\tau}_{\text{sc}}\lesssim1$ (see Appendix \ref{sefe} for a discussion). By combining Eq. (\ref{rmcri}), we find that the observational frequency is bounded within the range
\begin{equation}
	1.7\times10^{10}\tau_{\text{sc}}^* \text{RM}^{-1}\lesssim\nu_{\text{GHz}}\lesssim3.9\times10^{-4}\text{RM}^{1/3}W^{-1/3}.
	\label{frequencyrange}
\end{equation}
This suggests the necessity of a short pulse with width
\begin{equation}
	W\lesssim\left(10^{-41}\text{ s}\right)\tau_{\text{sc}}^{*-3}\text{RM}^4.
	\label{repulsewidth}
\end{equation}

Scattering measurements of FRB 121102A reveals that its broadening timescales scaled to 1 GHz vary from $\sim13\text{ }\mu\text{s}$ to $<0.89\text{ ms}$.\footnote{\cite{Michilli2018} and \cite{Hessels2019} indirectly estimated the scattering broadening to be $\sim13\text{ }\mu\text{s}$ and $\sim20\text{ }\mu\text{s}$ from scintillation, respectively. \cite{Josephy2019} and \cite{CHIME/FRBCollaboration2021} measured the scattering broadening to be $\sim0.27\text{ ms}$ and $<0.89\text{ ms}$, respectively.} If we adopt a RM value $\sim10^5\text{ rad m}^{-2}$, then the relative scattering broadening ranges from $\sim3$ to $\sim100$ at $1\text{ GHz}$. Even if the minimal scattering broadening $\tau_{\text{sc}}^{*}\sim13\text{ }\mu\text{s}$ is adopted, a frequency $\gtrsim2\text{ GHz}$ and a pulse width $\lesssim1\text{ }\mu\text{s}$ are still required. For some repeater sources with large RM values, much longer scattering timescales are expected \citep{Feng2022,Yang2022}. For example, the averaged scattering broadening of FRB 190520B scaled to 1 GHz can reach over $\sim10\text{ ms}$ \citep{Niu2022,Ocker2023}. This may suggest that detecting the extreme Faraday effect is more promising in sources like FRB 121102A, where scattering may not originate from the material responsible for most of its RM.

\section{Conclusions and Discussions}\label{cdiscussion}

A series of new features in an FRB were predicted when its RM exceeds the critical value given by Eq. (\ref{rmcri}). These are caused by the splitting of the R-mode and L-mode components after the FRB passes through a magnetized, sufficiently dense electron--ion plasma. Although the critical RM has not yet been reached by current observations, it may be exceeded in the future. Several observational challenges exist, with scattering being the most severe, which restricts the observation frequency to be relatively high and the pulse width to be extremely narrow. A similar phenomenon may also occur when a radio burst passing through the dense electron--positron pair plasma in the magnetosphere of its neutron star companion. In this case, an abrupt jump in the PA could be observed, despite a relatively low probability.

The two separated sub-bursts can be used to diagnose the plasma properties on the timescale comparable to the burst duration. Since the two sub-bursts originate from a same initial burst, it is expected to find highly consistent morphologies in both sub-bursts. In contrast, if these two twin bursts exhibit distinct morphologies from each other, it could suggest that the plasma properties on the propagation path vary over an extremely short timescale. It is worth noting that two ordinary adjacent bursts with short waiting times are sometimes not able to perform such a fine comparison, since it cannot be guaranteed that some of the morphological differences do not arise from their own radiation processes.

Very recently, it was argued by \cite{Wu2024} that the circular polarization of some isolated nanoshots in a giant pulse \citep{Hankins2003} arises from the splitting of original linearly polarized pulses through the extreme Faraday effect. We also note that some phenomena in FRBs resemble those predictions in this paper. For instance, the sudden orthogonal jump in PAs \citep{Niu2024} is very similar to the splitting of the X-mode and O-mode components after an FRB passing through a thick pair plasma. Some bursts exhibit noticeable sign-reversal behavior in circular polarization within an individual pulse \citep{Cho2020,Jiang2024}. However, explaining these phenomena by the extreme Faraday effect is not easy, as some other possible accompanying predictions were absent in the meanwhile, such as the clear signal of rapid Faraday conversion or large RM. Therefore, future high-precision observations are still needed to further confirm the existence of the extreme Faraday effects in FRBs.

\section{Acknowledgments}

We thank the referee very much for helpful comments that have allowed us to improve our manuscript significantly. The authors are very grateful to Su-Ming Weng for providing the presentation on the extreme Faraday effect. Y.C.H thanks Zhen-Yin Zhao and Sen-Lin Pang for useful discussions. This work was supported by  the National Natural Science Foundation of China (grant No. 12393812), the National SKA Program of China (grant No. 2020SKA0120302), and the Strategic Priority Research Program of the Chinese Academy of Sciences (grant NO. XDB0550300).

\appendix

\section{A General Discussion on Equations (3) and (14)}\label{gdes}

In some plasmas with weak dispersion (i.e., $v_p\simeq c \text{ and }  v_g\simeq c$), the phase difference between two permitted propagation modes (e.g., A-mode and B-mode) can usually be written as an unified form,
\begin{equation}
	\phi_A-\phi_B=M\lambda^\Gamma,
\end{equation}
where $M>0$ is a measurable quantity determined by the plasma, and the index $\Gamma$ is a constant. We conclude that the condition for the extreme Faraday effect in such a plasma can be equivalently written as
\begin{equation}
	\phi_A-\phi_B\gtrsim\frac{1}{\left|\Gamma\right|}\omega W,
\end{equation}
or alternatively,
\begin{equation}
	M\gtrsim M_{c}\equiv\frac{\omega^{\Gamma+1}W}{\left(2\pi c\right)^\Gamma \left|\Gamma\right|}.
\end{equation}
The constant index can be expressed by
\begin{equation}
	\Gamma=-\lim_{\text{wd}}\frac{\Delta v_g}{\Delta v_p},
\end{equation}
where the limit is taken under the weak dispersion approximation.

This conclusion provides a convenient method to determine whether the extreme Faraday effect occurs in a plasma. We provide some examples here to illustrate it.
\begin{itemize}
	
	\item In a weakly magnetized electron--ion plasma with $\omega_p\ll\omega$ and $\omega_B\ll\omega$, the ordinary Faraday effect occurs in an electromagnetic wave. One can easily obtain $\Gamma_{\text{ei}}=2$. The condition required for the extreme Faraday effect can be written as
	$\phi_R-\phi_L\gtrsim\omega W/2$.
	The critical value of $M_{\text{ei}}$ is given by $M_{\text{ei},c}=\omega^3 W/8\pi^2 c^2$. In fact, one can further verify that $M_{\text{ei}}=2\text{RM}$. Therefore, the critical $\text{RM}$ is consistent with Eq. (\ref{rmcri}).
	
	\item In a strongly magnetized electron--positron pair plasma with $\omega_p\ll\omega$ and $\omega\ll\omega_B$, one has $\Gamma_{\text{ep}}=1$. The condition required for the extreme Faraday effect is $\phi_X-\phi_O\gtrsim\omega W$, or equivalently, $M_{\text{ep},c}=\omega^2 W/2\pi c$. One can further verify that $M_{\text{ep}}=\left(2e^2/m_e c^2\right)\text{DM}_{\text{pair}} $. Therefore, the derived critical $\text{DM}_{\text{pair}}$ is consistent with Eq. (\ref{epnedl}).
	
	\item In a weakly magnetized electron--positron pair plasma with $\omega_p\ll\omega$ and $\omega_B\ll\omega$, one has $\Gamma_{\text{ep}}^*=3$. The condition for the extreme Faraday effect is $\phi_O-\phi_X\gtrsim\omega W/3$, or $M_{\text{ep},c}^{*}=\omega^4 W /24\pi^3 c^3$, where the parameter $M_{\text{ep}}^*=\left(e^4/2\pi^2 m_e^3 c^6\right)\int n_e B_{\parallel}^2 dl$. This gives $\int n_e B_{\parallel}^2 dl\gtrsim\left(5.2\times10^{15}\text{ pc cm}^{-3}\text{ }\mu\text{G}^2\right)\nu_\text{GHz}^4 W_{-3}$.
	
	\item In a strongly magnetized electron--ion plasma with $\omega_p\ll\omega$ and $\omega\ll\omega_B$, one has $\Gamma_{\text{ei}}^*=0$. This result indicates that the extreme Faraday effect is almost impossible to occur in such a plasma.
\end{itemize}
Based on the classical magnetized plasma dispersion equation, we conclude that the extreme Faraday effect is unlikely to occur in a weakly magnetized pair plasma or a strongly magnetized electron--ion plasma, which do not receive much attention in this paper.

\section{Depolarization due to Multipath Propagation in the Extreme Faraday Effect}\label{depefe}

We suppose that the pulse profiles of the L-mode and the R-mode follow simple Gaussian distributions, i.e.,
\begin{equation}
	I_L=\text{exp}\left\{-\frac{\left(t+\Delta T/2\right)^2}{2\sigma_t^2}\right\},\text{ }I_R=\text{exp}\left\{-\frac{\left(t-\Delta T/2\right)^2}{2\sigma_t^2}\right\},
	\label{pulprolr}
\end{equation}
where $\Delta T=\left(5.7\times10^{-11}\text{ s}\right)\nu_{\text{GHz}}^{-3}\text{RM}$ is the interval between the two sub-pulses given by Eq. (\ref{deltat}), and $\sigma_t$ is a parameter related to the pulse width. We also neglect the constant coefficients since both of them can be normalized finally. The Stokes parameters of the superposed wave are then given by
\begin{equation}
	Q\left(t,\Delta T,\phi\right)=2\sqrt{I_L I_R}\cos{\left(2\phi\right)},\text{ }U\left(t,\Delta T,\phi\right)=2\sqrt{I_L I_R}\sin{\left(2\phi\right)},\text{ }I\left(t,\Delta T\right)=\left(I_L+I_R\right),\text{ }V\left(t,\Delta T\right)=-\left(I_L-I_R\right),
\end{equation}
where $\phi=\text{RM}\lambda^2$ is the PA. Multipath propagation indicates that rays going through different paths possess different RM values. A reasonable assumption is that their RM values are governed by a Gaussian distribution,
\begin{equation}
	\mathcal{N}\left(\text{RM}_0,\sigma_{\text{RM}}^2\right)=\frac{1}{\sqrt{2\pi}\sigma_{\text{RM}}}\text{exp}\left\{-\frac{\left(\text{RM}-\text{RM}_0\right)^2}{2\sigma_{\text{RM}}^2}\right\}.
\end{equation}
The distribution can modulate the total observed Stokes parameters due to the superposition of different rays, i.e.,
\begin{equation}
	\begin{aligned}
		Q_{\text{tot}}&=\int Q\left(t,\Delta T,\phi\right)\mathcal{N}\left(\text{RM}_0,\sigma_{\text{RM}}^2\right)d\text{RM},\text{ }
		U_{\text{tot}}=\int U\left(t,\Delta T,\phi\right)\mathcal{N}\left(\text{RM}_0,\sigma_{\text{RM}}^2\right)d\text{RM},\\
		I_{\text{tot}}&=\int I\left(t,\Delta T\right)\mathcal{N}\left(\text{RM}_0,\sigma_{\text{RM}}^2\right)d\text{RM},
		\text{ }
		V_{\text{tot}}=\int V\left(t,\Delta T\right)\mathcal{N}\left(\text{RM}_0,\sigma_{\text{RM}}^2\right)d\text{RM}.
	\end{aligned}
\end{equation}
These integrals are complex since both $\Delta T$ and $\phi$ involve the $\text{RM}$ term. However, they can be simplified by assuming $\sigma_{\text{RM}}\ll\text{RM}_0$, which is supported by observations \citep{Feng2022}. As a result, the difference of the interval between the L-mode and R-mode of any ray caused by the RM scatter can be neglected. We can then treat $\Delta T$ as a constant and extract it from the integral.\footnote{Strictly speaking, the interval difference due to RM scatter can be neglected only when compared to $\Delta T(\text{RM}_0)$. Therefore, $\Delta T$ can be extracted from the integral when $\Delta T(\text{RM}_0)\ll\sigma_t$ or $\Delta T(\text{RM}_0)\sim\sigma_t$, either of which is reasonable when the extreme Faraday effect is not exceptionally strong.} The final Stokes parameters thus are
\begin{equation}
	\begin{aligned}
		Q_{\text{tot}}&\simeq2\sqrt{I_L I_R}\cos{\left(2\text{RM}_0\lambda^2\right)}\text{exp}\left\{-2\lambda^4 \sigma_{\text{RM}}^2\right\},\text{ }
		U_{\text{tot}}\simeq2\sqrt{I_L I_R}\sin{\left(2\text{RM}_0\lambda^2\right)}\text{exp}\left\{-2\lambda^4 \sigma_{\text{RM}}^2\right\},\\
		I_{\text{tot}}&\simeq\left(I_L+I_R\right),\text{ }V_{\text{tot}}\simeq-\left(I_L-I_R\right),\text{ }L_{\text{tot}}\simeq2\sqrt{I_L I_R}\text{ exp}\left\{-2\lambda^4 \sigma_{\text{RM}}^2\right\}.
	\end{aligned}
\end{equation}
One can see that the result does not depend on the concrete form of the pulse profile given by Eq. (\ref{pulprolr}). This result also indicates that the degree of linear polarization goes through depolarization, while the degree of circular polarization remains nearly unchanged after multipath propagation. When $I_L=I_R\neq0$, the degree of depolarization reduces to the case reported by \cite{Feng2022}, i.e., $f_{\text{dep}}=1-\text{exp}\left\{-2\lambda^4 \sigma_{\text{RM}}^2\right\}$. When either $I_L$ or $I_R$ vanishes, the degree of circular polarization equals to $100\%$, which corresponds the state of pure circular polarization in the extreme Faraday effect.

\section{Visualization of Scattering-Broadened Pulse Profiles}\label{sefe}

\begin{figure*}
	\begin{center}
		\subfigure{
			\includegraphics[width=0.23\textwidth]{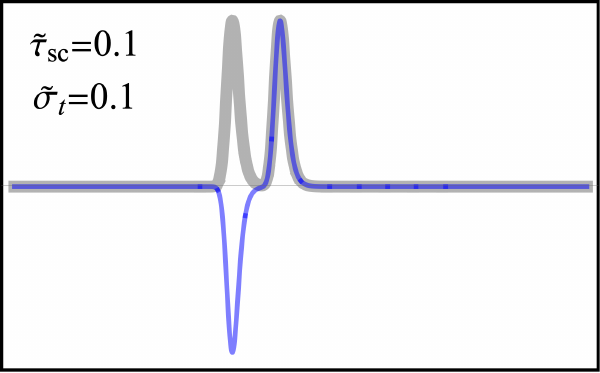}}
		\subfigure{
			\includegraphics[width=0.23\textwidth]{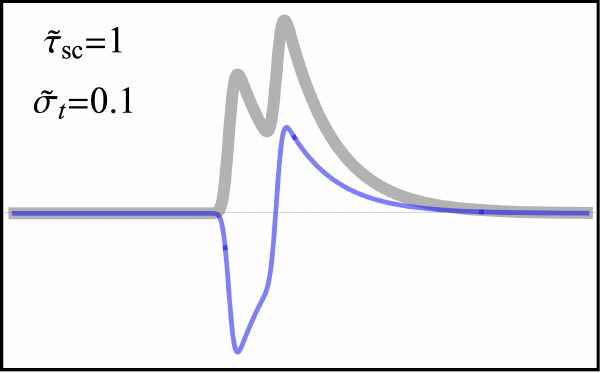}}
		\subfigure{
			\includegraphics[width=0.23\textwidth]{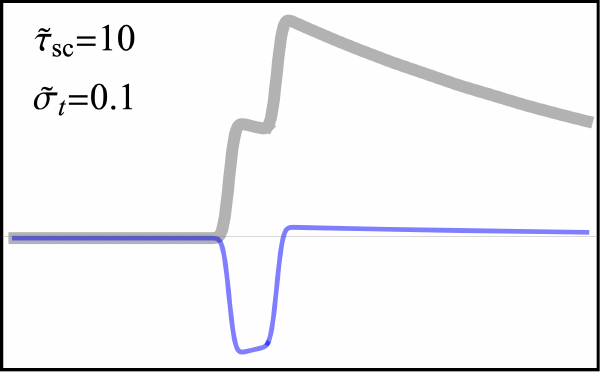}}
		\subfigure{
			\includegraphics[width=0.23\textwidth]{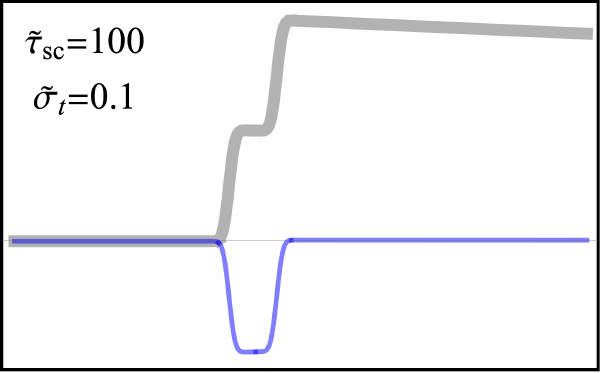}}
		\subfigure{
			\includegraphics[width=0.23\textwidth]{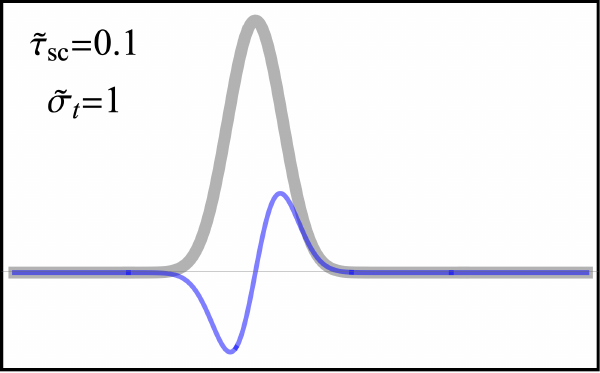}}
		\subfigure{
			\includegraphics[width=0.23\textwidth]{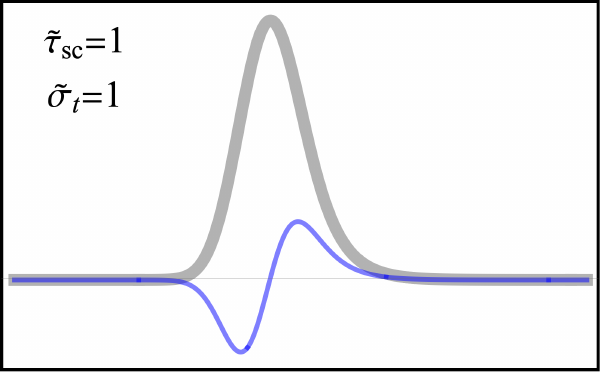}}
			\subfigure{
			\includegraphics[width=0.23\textwidth]{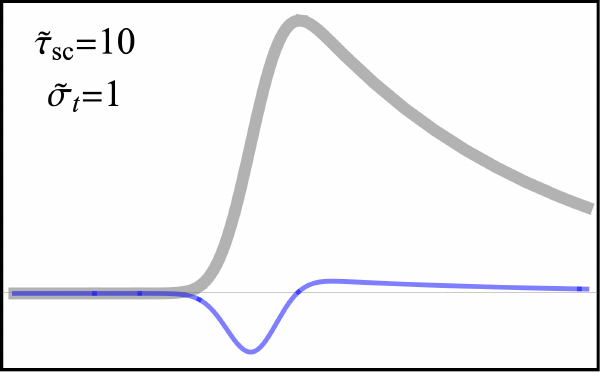}}
		\subfigure{
			\includegraphics[width=0.23\textwidth]{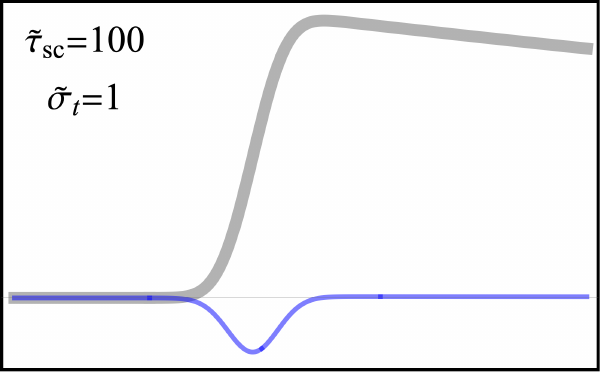}}
		\caption{Various pulse profiles with different relative scattering broadening $\widetilde{\tau}_{\text{sc}}$ and relative intrinsic width $\widetilde{\sigma}_t$. The gray thick curve and the blue curve represent the total intensity and the circularly polarized intensity, respectively. Scattering begins to smear one of the circularly polarized modes when $\widetilde{\tau}_{\text{sc}}\gtrsim1$. For an extremely large relative scattering broadening, the curve of circular polarization exhibits a dip structure, which is consistent with the case discussed by \cite{Suresh2019}. The linear polarization component may be depolarized and thus is not considered here.}
		\label{sbplot}
	\end{center}
\end{figure*}

In this appendix, we visualize the shape of the pulse with scattering broadening under the extreme Faraday effect. We do not require high accuracy and thus use the exponentially modified Gaussian (EMG) distribution, which is defined as the convolution of a Gaussian and an exponential function, to approximate the pulse profile. The EMG distributions of the L-mode and R-mode have the following forms,
\begin{equation}
	\begin{aligned}
		I_L&=\frac{1}{2\tau_{\text{sc}}}\text{exp}\left\{\frac{1}{\tau_{\text{sc}}}\left[-\frac{\Delta T}{2}+\frac{\sigma_t^2}{2\tau_{\text{sc}}}-t\right]\right\}\text{erfc}\left\{\frac{-\Delta T/2+\sigma_t^2/\tau_{\text{sc}}-t}{\sqrt{2}\sigma_t}\right\},\\
		I_R&=\frac{1}{2\tau_{\text{sc}}}\text{exp}\left\{\frac{1}{\tau_{\text{sc}}}\left[\frac{\Delta T}{2}+\frac{\sigma_t^2}{2\tau_{\text{sc}}}-t\right]\right\}\text{erfc}\left\{\frac{\Delta T/2+\sigma_t^2/\tau_{\text{sc}}-t}{\sqrt{2}\sigma_t}\right\},
	\end{aligned}
\end{equation}
where $\tau_{\text{sc}}$ is the timescale of scattering broadening, and $\text{erfc}(x)=\left(2/\sqrt{\pi}\right)\int_x^{\infty}e^{-t^2}dt$ is the complementary error function. When $\tau_{\text{sc}}\to0$, the above EMG distributions reduce to ordinary Gaussian distributions $\mathcal{N}\left(\mp\Delta T/2,\sigma_t^2\right)$. The pulse shape is completely determined by two parameters, i.e., the relative scattering broadening $\widetilde{\tau}_{\text{sc}}=\tau_{\text{sc}}/\Delta T$, which characterizes the degree of scattering, and the relative intrinsic width $\widetilde{\sigma}_t=\sigma_t/\Delta T$, which characterizes the separation between the two sub-pulses. Various pulse profiles with different $\widetilde{\tau}_{\text{sc}}$ and $\widetilde{\sigma}_t$ values are shown in Fig. \ref{sbplot}. The sign-switching behavior of circular polarization begins to be smeared when $\widetilde{\tau}_{\text{sc}}\gtrsim1$. In particular, when $\widetilde{\tau}_{\text{sc}}$ is extremely large, the circular polarization profile exhibits a dip structure, which returns to the case discussed by \cite{Suresh2019}.

\normalem
\bibliography{EFR}

\end{document}